\documentclass[10pt,twocolumn]{article}
\usepackage{graphicx}

\usepackage{authblk}
\usepackage{float}
\usepackage{braket}
\usepackage{mathtools}
\usepackage{array}
\usepackage[acronym]{glossaries}
\newtheorem{definition}{Definition}

\usepackage{amsmath}
\usepackage{mathrsfs}
\usepackage{xcolor}
\usepackage{amssymb}
\usepackage{verbatim}
\usepackage{mathabx}
\usepackage{bbold}
\usepackage[font=small,labelfont=bf]{caption}
\usepackage[super]{natbib}
\usepackage[hidelinks]{hyperref}
\usepackage[normalem]{ulem}
\usepackage[linesnumbered,ruled]{algorithm2e}
\usepackage[titletoc,toc,title]{appendix}
\usepackage{chemfig}
\usepackage{graphicx}
\usepackage{caption}
\usepackage{subcaption}

\makeglossaries

\title{Quantum information theory on sparse wavefunctions and applications for Quantum Chemistry}

\author[1,2]{Davide Materia}
\author[1,3] {Leonardo Ratini}
\author[1]{Leonardo Guidoni
\thanks{leonardo.guidoni@univaq.it}}

\affil[1]{
Dipartimento di Scienze Fisiche e Chimiche, Universit\`a degli Studi dell’Aquila, Coppito, L’Aquila, Italy }

\affil[2]{Dipartimento di Ingegneria e Scienze dell'Informazione e Matematica\\ Universit\`a degli Studi dell'Aquila, Coppito, L'Aquila, Italy}

\affil[3]{Dipartimento di Scienze Matematiche, Fisiche e Informatiche\\ Universit\`a degli Studi di Parma, Parma, Italy}

\date{\today}

\begin{document}

\newacronym{qc}{QC}{Quantum Computing}
\newacronym{sparq}{SparQ}{Sparse Quantum State Analysis}
\newacronym{mps}{MPS}{Matrix Product State}
\newacronym{dmrg}{DMRG}{Density Matrix Renormalization Group}
\newacronym{dm}{DM}{Density Matrix}
\newacronym{hf}{HF}{Hartree-Fock}
\newacronym{mp2}{MP2}{Møller-Plesset of second order}
\newacronym{ci}{CI}{Configuration Interaction}
\newacronym{mrci}{MRCI}{Multi-Reference Configuration Interaction}
\newacronym{cisd}{CISD}{CI with singles and doubles}
\newacronym{cc}{CC}{Coupled Cluster}
\newacronym{ccsd}{CCSD}{CC with singles and doubles}
\newacronym{scf}{SCF}{Self Consistent Fields}
\newacronym{sd}{SD}{Slater Determinants}
\newacronym{mo}{MO}{Molecular Orbitals}
\newacronym{so}{SO}{Spin Orbitals}
\newacronym{hfco}{HFCO}{Hartree-Fock Canonical Orbitals}
\newacronym{no}{NO}{Natural Orbitals}
\newacronym{ino}{INO}{Iterative Natural Orbitals}
\newacronym{noon}{NOON}{Natural Orbitals Occupation Number}
\newacronym{qmi}{QMI}{Quantum Mutual Information}
\newacronym{mi}{MI}{Mutual Information}
\newacronym{tpcpm}{TPCPM}{Trace Preserving Completely Positive Map}
\newacronym{cas}{CAS}{Complete Active Space}
\newacronym{mrpt2}{MRPT2}{Multi-Reference Perturbation Theory}
\newacronym{fci}{FCI}{Full CI}
\newacronym{ivo}{IVO}{Improved Virtual Orbital}
\newacronym{mvo}{MVO}{Modified Virtual Orbital}
\newacronym{icscf}{ICSCF}{Internally Consistent SCF}

\twocolumn[
\begin{@twocolumnfalse}
\maketitle
\begin{abstract}

In recent years Quantum Computing prominently entered in the field of Computational Chemistry, importing and transforming computational methods and ideas originally developed within other disciplines, such as Physics, Mathematics and Computer Science into algorithms able to estimate quantum properties of atoms and molecules on present and future quantum devices. 
An important role in this contamination process is attributed to Quantum Information techniques, having the twofold role of contributing to the analysis of electron correlation and entanglements and guiding the construction of wavefunction variational ansatzes for the Variational Quantum Eigensolver technique.

This paper introduces the tool SparQ (Sparse Quantum state analysis), designed to efficiently 
compute fundamental quantum information theory observables on post-Hartree-Fock wavefunctions sparse in their definition space. The core methodology involves mapping fermionic wavefunctions to qubit space using fermionic-to-qubits transformations and leveraging the sparse nature of these wavefunctions to evaluate observables and properties of the wavefunction.

The effectiveness of SparQ is validated by analyzing the mutual information matrices of wavefunctions for the water molecule and the total entropy of $\sim 10^2$ qubits describing the benzene moleculehighlights. This way its capability to handle large-scale quantum systems, limited mainly by the capabilities of quantum chemical methods used to retrieve the wavefunctions.
The results indicate that quantum information theoretical analysis, so far limited to traditional tensor network methods and study of transition operators, can be applied to all post-Hartree-Fock wavefuncions, extending their applications to larger and more complex chemical systems. 
\vspace{3cm}

\end{abstract}
\vspace{1cm}
\end{@twocolumnfalse}

]

\newpage

\section{\label{sec:intro}Introduction}

In recent years, the rapid growth of \acrfull{qc}\cite{physchem} has highlighted the relevance of quantum information theory tools within computational sciences.~\cite{Feynman1982,Shor1997,Banuls2020,Hussain2020} 
This convergence has promoted a reciprocal exchange of ideas between computational sciences and \acrshort{qc}, moving us closer to the era of hybrid quantum-classical computing.

In the field of Quantum Chemistry~\cite{Hartree1928, Szabo2012}, this exchange has enabled for existing algorithms\cite{Aspuru-Guzik2005, Grimsley2019, Ratini2022, Ratini2023,Egger2023} to benefit from the inclusion of chemical concepts, exemplified by the UCCSD ansatz~\cite{Hoffmann1998, Kutzelnigg1991, Cooper2010, Evangelista2011, Whitfield2011} for variational quantum chemistry algorithms. Additionally, advancements in quantum information theory have benefited classical quantum chemistry by enhancing the understanding of chemical properties~\cite{barcza2011, Boguslawski2015, Ding2021} and introducing tools that focus computational efforts where they are most needed.~\cite{Stein2016}

Despite increasing interest in quantum information theory analysis within computational sciences, most existing works rely on Tensor Network methods, which facilitate the intuitive handling of wavefunctions to extract quantum information properties. Within this framework, the \acrfull{dmrg}\cite{White1992, White93, Legeza2003, Schollwock2005, SCHOLLWOCK2011} algorithm has proven to be a highly capable and versatile tool~\cite{Boguslawski2014, bensberg22}, retrieving high-quality wavefunctions in a description, which can then be analyzed using various techniques to extract quantum information.  \\
However, certain applications of tensor network methods exhibit significant limitations in system size and quality. In quantum chemistry, the non-locality of the electronic Hamiltonian makes \acrshort{dmrg} calculations for systems larger than a few hundred qubits impractical, leaving a gap for quantum information theoretical analysis on larger systems.

This work tries to fill this gap by introducing the \acrfull{sparq} tool. We here propose a method to effectively calculate the most basic tools of quantum information theory for any wavefunction that is sparse in its definition space. The main focus of this procedure is to manipulate quantum chemistry's wavefunctions, coming from Post-\acrfull{hf} methods, directly to the qubit space through a fermion-to-qubit mapping~\cite{Jordan1928, Miller2023}.
Furthermore, given the inspiration from which this work arises\cite{Materia2024}, it falls directly in line with the mentioned concept of bringing \acrshort{qc} and quantum computing closer.

The manuscript is organized as follows, at the beginning we will dive deeper into the necessity of a tool different from tensor networks to analyze wavefunctions, justifying it in part with the chemical background that has sparked the idea for this work. Following this track, we will show how to encode fermionic wavefunction to the qubit space.\\
We then proceed with a detailed explanation of the developed tools, with an analysis of the performance and computational costs of the introduced procedures. We conclude by proposing potential usage of the method and by showing practical applications. 

\section{\label{sec:qc}Wavefunction representations and sparsity in Quantum Chemistry methods}
Given the focus of the authors and the method's primary application, this section is dedicated to giving a context to the use of wavefunctions in quantum chemistry and to show how one can map them to the qubit space. However, since this is not essential for understanding the quantum information part of this work, we invite a reader solely interested in analyzing sparse wavefunctions in qubit space to divert over section \ref{sec:sparq} and forward.
In quantum chemistry, the \acrshort{mps}-based \acrshort{dmrg}\cite{SCHOLLWOCK2011} has been the only method used for quantum information theoretical analysis. However, the \acrshort{dmrg} algorithm cannot handle quantum chemistry applications involving more than a few hundred qubits, which correspond (roughly) to the same amount of spin orbitals. This is due to the scaling of the \acrshort{dmrg} algorithm depending heavily on the so-called \textit{bond dimension} $\chi$, with the scaling being lower bounded by $O(N^3\chi^2)+O(N^2\chi^3)$~\cite{Wouters2014}. This scaling, even if particularly appealing compared to other methods for small to medium sized systems, quickly becomes unmanageable, especially given the growing requirements in terms of $\chi$ with growing system sizes.

By contrast, many other Post-\acrshort{hf} methods, given their excitation-based structure, achieve a total scaling as low as $O(N^5)$ (for \acrfull{mp2}\cite{Moller1934}), which renders possible calculations on up to thousands of spin orbitals ($\sim$qubits). Clearly, such a precision-oriented algorithm as \acrshort{dmrg} should not be so unfairly compared to perturbation theory (or other low-quality) methods, however, the possibility of generalizing quantum information analysis by using wavefunctions coming from other methods could extend its reach to previously unfeasible systems. \\ For completeness, we remark that other techniques, based on transition operators,~\cite{Rissler2006, Boguslawski2013, Boguslawski2015} also apply to methods different than \acrshort{dmrg}, since the necessary ingredients are only some terms of the n-body operators. However, besides the limited application that they find within other Post-\acrshort{hf} methods, a transition-operator-based study can only be undergone for a few qubits, given the complexity of the derivation of these operators. 

The stated generalizing purpose raises the point of bringing fermionic wavefunctions defined on the Fock space ($\mathcal{F}$) to the separable space of the qubits. This conceptually non-trivial task is handled by the next subsection. For what concerns us for the moment, the wavefunctions ($\ket{\psi}$) under scope can always be written as follows:

\begin{equation}
    \ket{\psi}_{\mathcal{F}} = \sum_{\mathbf{i}=\{0,1\}^N} c_{\mathbf{i}} \ket{\mathbf{i}}_{\mathcal{F}}
    \label{eq:ciwf}
\end{equation}
where each $\ket{\mathbf{i}}$ is a \acrshort{sd}, a vector of a basis of $\mathcal{F}$.

By this definition, there could be a number of components in the wavefunction growing exponentially with $N$, the number of \acrfull{so}. However, most Post-\acrshort{hf} methods such as \acrshort{mp2}~\cite{Moller1934}, \acrfull{ci}~\cite{Szabo2012}, \acrfull{cc}, and their multireference variants~\cite{neese2007, Sivalingam2016,caspt2,nevpt2}, employ relatively few states out of the exponentially many in \eqref{eq:ciwf}, this is because they usually 
 rely only on a limited number of excitations. 
 For example, in the case of the common double excitations limit, the expansion in the wavefunction of eq~\eqref{eq:ciwf} grows as the number of possible double excitations, which scales as $O(n_{o}^2n_{v}^2)$, for $n_{o}$ representing the occupied orbitals and $n_{v}$ the unoccupied ones. 
The actual space is then reduced to a polynomial number of \acrshort{sd} when the total space is still of exponential dimension.

\subsection{\label{subsec:encoding} Encoding of fermionic wavefunctions}

Beginning with the wavefunction definition in \eqref{eq:ciwf}, we demonstrate how to map the corresponding fermionic excitations to Pauli operators in the separable qubits space using a fermionic-to-qubit mapping.

To maintain a consistent notation for both Pauli operators, \acrshort{sd} and \eqref{eq:ciwf}, $\mathbf{i}$  will represent the binary string with ones indicating occupied fermionic modes. This means that, for a system of $N$ fermionic modes,
\begin{equation}
    a_{\mathbf{i}}^{\dag}=\prod_{j=1}^N \left( a_j^{\dag}\right)^{i_j} \qquad 
   {i_j}=0,1 \qquad j=1,\dots,N
\end{equation}
Consequently, the operator $a_{\mathbf{i}}^{\dag}$ applies the ordered creation operators of the ones in the $\mathbf{i}$ binary string.
In this framework, every state belonging to the Fock space can be represented as  
\begin{equation}
    \ket{\psi} = \sum_{\mathbf{i}=\{0,1\}^N} c_{\mathbf{i}} a_{\mathbf{i}}^{\dag} \ket{\varnothing}
    \label{eq:dagtonull}
\end{equation}
where $\ket{\varnothing}$ is the vacuum state.
As an example, for a system described with 4 spin orbitals, the string $\mathbf{i} = \{1010\}$ is associated to the following Slater determinant $a_{\mathbf{\{ 1010 \}}}^{\dag} \ket{\varnothing}= a_{1}^{\dag} a_{3}^{\dag} \ket{\varnothing} =\ket{1010}$.

The Fock space generated by $N$ fermionic modes can be mapped to the $N$-partite Hilbert space(the qubits space) by means of a \textit{fermionic-to-qubit} mapping, referred to as $\mathcal{K}$. The mapping associates to each creation operator $a_{j}^{\dag}$ the corresponding operator in the qubits space $a_{q,j}^{\dag}$ for each $j=1,\dots,N$. 
The mapping preserves the Fermi algebra associated to the fermionic modes

\begin{equation}
    \label{eq:comrel}
    \begin{split}
        &\{ a_{i},a_{j}\}=\{ a_{q,i},a_{q,j}\}\\
        &\{ a^{\dag}_{i},a^{\dag}_{j}\}=\{ a^{\dag}_{q,i},a^{\dag}_{q,j}\}\\
        &\{ a_{i},a^{\dag}_{j}\}=\{ a_{q,i},a^{\dag}_{q,j}\}=\delta_{i,j}\mathbb{1}
    \end{split}
\end{equation}

In \cite{Miller2023} it is shown how there exists a class of fermionic-to-qubit mappings such that the vacuum state $\ket{\varnothing}$ is mapped to the vector of the computational basis with all zeros $\ket{\mathbf{0}}$.  Assuming that the used $\mathcal{K}$ belongs to this class, we summarize the action of the mapping as follows 

\begin{align}
     & \mathcal{K}(\ket{\varnothing}) = \ket{\mathbf{0}} \\
& \mathcal{K}(a_{j}^{\dag}) = a_{q,j}^{\dag} =\frac{1}{2}(\gamma_{q}^{2j-1}-i\gamma_{q}^{2j}) \\
& \mathcal{K}(a_{j}) = a_{q,j} =\frac{1}{2}(\gamma_{q}^{2j-1}+i\gamma_{q}^{2j}) 
\label{eq:mapping} 
\end{align}
where $\gamma_{q}^{l}$ are the Majorana strings, i.e. anticommuting operators in the qubit space associated to the Majorana fermions defined in the Fock space.
Now, given this framework, we have Majorana operators of the following form:

\begin{equation}
    \gamma_{q}^{l}=\bigotimes_{m=1}^N \sigma_{l,m}
\end{equation}

where $\sigma_{l,m}=X_m,Y_m,Z_m,I_m$ are the Pauli matrices with the identity, acting on qubit $m=1,\dots,N$.

To each creation operator we can associate the value of the Hamming distance calculated taking into account the associated Majorana strings
\begin{equation}
    \label{eq:hamming_distance}
    \mathcal{H}(a_{q,j}^{\dag})=\mathcal{H}(\gamma_{q}^{2j}, \gamma_{q}^{2j-1}) = \sum_{m=1}^{N} \delta_{\sigma_{2j,m},\sigma_{2j-1,m}}
\end{equation}
Suppose that for each of the $N$ fermionic modes the Hamming distance is greater than one. In this general case, a product of $P$ fermionic modes is required to store $2^P$ Majorana strings to encode the wavefunction. This is the case of the parity mapping, as can be easily understood by considering the corresponding tree shown in \cite{Miller2023}. However, generally, this prohibitive cost can be reduced if just a subset of the maiorana pairs associated to a given excitation operator has Hamming distance greater than 1. More information regarding the procedure for such cases is given in Appendix \ref{appendix:mapgeneral}. 

\subsection{Jordan Wigner mapping}
The Jordan-Wigner\cite{Jordan1928} mapping, one of the most used and ancient mappings, has Hamming distance equal to one for any mode. This fact renders particularly convenient the encoding of a wavefunction in this mapping. Let us first of all show its definition:

\begin{equation}
    \sigma_{l,m}=
        \begin{cases}
            I_m & \text{if } \lfloor \frac{l}{2} \rfloor>m \\
            X_m & \text{if } \lfloor \frac{l+1}{2} \rfloor=m \\
            Y_m & \text{if } \lfloor \frac{l}{2} \rfloor=m\\
            Z_m & \lfloor \frac{l}{2} \rfloor-m
        \end{cases}
    \label{eq:jw}
\end{equation}

Using this definition, Eq.~\eqref{eq:mapping} assumes the following form:

\begin{equation}
    \begin{split}
        & a_{q,j}^{\dag} = \bigotimes_{m=1}^{N} \epsilon_{j,m} \quad with \\
        &\epsilon_{j,m} = 
        \begin{cases}
            \sigma_{2j,m},& \text{if } \sigma_{2j,m}=\sigma_{2j+1,m} \\
            \frac{1}{2} (\sigma_{2j,m}-i\sigma_{2j+1,m}), & \text{otherwise}
        \end{cases}
    \end{split}
    \label{eq:contraction}
\end{equation}
This allows us to rewrite a single SD of Eq.~\eqref{eq:dagtonull} as 

\begin{equation}
    \mathcal{K}(a_{\mathbf{i}}^{\dag}) = a_{q,\mathbf{i}}^{\dag} = \bigotimes_{m=1}^{N}  \prod_{l=1}^N (\epsilon_{l,m})^{\mathbf{i}_l}
    \label{eq:qubitSD}
\end{equation}

We can now rewrite directly Eq.~\eqref{eq:dagtonull} as 

\begin{equation}
\begin{split}
    &\ket{\psi} = \sum_{\mathbf{i}=\{0,1\}^N}c_{\mathbf{i}} a_{q,\mathbf{i}}^{\dag} \ket{\mathbf{0}} = \\
    &=\sum_{\mathbf{i}=\{0,1\}^N} c_{\mathbf{i}}  \left( \bigotimes_{m=1}^{N}  \prod_{l=1}^N (\epsilon_{l,m})^{\mathbf{i}_l}\right)  \ket{\mathbf{0}} 
\end{split}
    \label{eq:qubitdagtonull}
\end{equation}
Now, to pave the way for the usage of wavefunctions of this form in the following section~\ref{sec:sparq}, we write the wavefunction as follow

\begin{equation}
\begin{split}
    &\ket{\psi}=\sum_{\mathbf{i}=\{0,1\}^N} c_{\mathbf{i}}  \left( \bigotimes_{m=1}^{N} \ket{\psi}_{\mathbf{i},m}\right) = \sum_{\mathbf{i}=\{0,1\}^N} c_{\mathbf{i}} \ket{\psi}_{\mathbf{i}}
\end{split}
\label{eq:absorption}
\end{equation}
 where  
 
\begin{equation}
   \ket{\psi}_{\mathbf{i},m}=\epsilon_{\mathbf{i},m}\ket{0}_m, \quad \epsilon_{\mathbf{i},m} = \prod_{l=1}^N (\epsilon_{l,m})^{\mathbf{i}_l}
    \label{eq:psiim}
\end{equation}

\begin{equation}                    
and \quad \ket{\psi}_{\mathbf{i}}=\bigotimes_{m=1}^{N} \ket{\psi}_{\mathbf{i},m}
    \label{eq:psii}
\end{equation}

our ket is now written as a sum of tensor products of single-qubit wavefunctions $\ket{\psi}_{i,m}$.
The Eq.~\eqref{eq:absorption} is the final expression for a generic wavefunction represented on a qubit space, where the sum runs over all the possible \acrshort{sd}. However, given the focus of the present work on Post-Hartree Fock \acrshort{ci}-based methods, this sum only runs over the possible excitations on a reference wave function. 

One disclaimer, for the simple case of the Jordan Wigner mapping, one could also map the wavefunction by hand, avoiding this lengthy procedure, as the $a^\dag$ operators for this mapping are quite simple and translate directly Eq.~\eqref{eq:ciwf}
 to a binary expression over the qubit space, so that the following is true:

\begin{equation}
    \sum_{\mathbf{i}} c_{\mathbf{i}} \ket{\psi}_{\mathbf{i}} = \sum_{\mathbf{i}} k_{\mathbf{i}}\ket{\mathbf{i}}
    \label{eq:qubitkets}
\end{equation}
where $\ket{\mathbf{i}}$ is binary ket.
On the other hand, this is equivalent to realizing that in \eqref{eq:contraction} and \eqref{eq:psiim} no superposition is created by any $\epsilon_{l,m}\ket{0}_m$, as it can be seen by noting that:
\begin{equation}
    \epsilon_{\mathbf{i},m}\ket{0}_m =
    \begin{cases}
            \ket{0}_m & \text{if } \mathbf{i}_m = 0 \\
            \pm \ket{1}_m & \text{if } \mathbf{i}_m = 1
     \end{cases}
\end{equation}
So, for each $m$ we never obtain a linear combination of the two states. This will be used in the following to reduce the computational cost of the trace operator.

The long process for mapping the wavefunction must then be considered for more elaborate mappings, which are built upon the technique explained in this section (Appendix \ref{appendix:mapgeneral}).

\section{\label{sec:qit}Quantum Information}

The primary objective of this work is to develop an efficient method for computing information-theoretical quantities for non-\acrshort{mps} wavefunctions. To achieve this, we must first evaluate the general complexity of such analyses.

In quantum information theory, analyses typically occur in a Hilbert space $\mathscr{H}$, where each vector represents a wavefunction $\ket{\psi}_{\mathscr{H}}$. 
In quantum computing, this Hilbert space consists of multiple separable two-dimensional spaces known as \textit{qubits}. The entire space of $N$ qubits $\mathscr{H}$ is
\begin{equation}
    \mathscr{H}=\bigotimes_{l}^{N}\mathscr{H}_l
    \label{eq:tensordot}
\end{equation}
with a total dimension of $\mathscr{H}$ is $2^N$ \cite{Nielsen2010}.

Given a wavefunction $\ket{\phi} \in \mathscr{H}$, its \acrfull{dm} can defined over the space of \textit{Endomorphisms} of $\mathscr{H}$, $End(\mathscr{H})$, such that 
\begin{equation}
    \label{eq:dm}
    \rho_{\phi}=\ket{\phi}\bra{\phi}
\end{equation}
This state description captures general quantum or statistical properties, but it requires a quadratic cost in terms of the components of the wavefunction $\phi$.
Given the exponential dimension of the space in relation to the number of qubits, analyzing the properties of wavefunctions and density matrices within this space is impractical.

To reach the goal, there is a need to make use of the sparse property of the wavefunctions of chemical systems discussed in section~\ref{sec:qc}.

For a bipartite quantum system with Hilbert spaces $\mathscr{H}_A$ and $\mathscr{H}_B$, the partial trace over subsystem $B$ of the density matrix $\rho_{AB}$ is defined as:

\begin{equation}    
\rho_A = \text{Tr}(\rho_{AB})_{B}=\sum_{i} \prescript{}{B} {\bra{i}}\rho_{AB}\ket{i}_{B}
\label{eq:partial_trace}
\end{equation}

Where $\ket{i}_{B}$ are any orthonormal basis of the Hilbert space B.  The primary use of the traced density matrix is to calculate the Von Neumann (or quantum) mutual information.\cite{vonneumann1996}

\begin{definition}
Consider two Hilbert spaces $\mathscr{H}_A$ and $\mathscr{H}_B$, and let $\rho_{AB}\in \mathscr{H}_A \otimes \mathscr{H}_B$ be a density matrix. The quantum mutual information $I$ of $\rho_{AB}$ is given by:
\begin{equation}
    I(A,B) = S(A) + S(B) - S(A,B)
    \label{eq:qmi}
\end{equation}
where $S$ represents the von Neumann entropy defined as
\begin{equation}
\begin{split}
    S(A,B) &= -\operatorname{tr}(\rho_{AB}\log(\rho_{AB}))\\
    S(A) &= -\operatorname{tr}(\rho_{A}\log(\rho_{A}))\\
    \rho_{A} &= \operatorname{tr}_{B}(\rho_{AB})
\end{split}
\label{eq:entropy}
\end{equation}
\end{definition}

As described in Eq.~\eqref{eq:entropy}, for an N-partite Hilbert space $\mathscr{H}_N$ and a quantum state expressed by $\rho \in \mathscr{H}_N$, the elements $I(i,j)$ of the mutual information matrix $I$ are obtained by tracing out all subsystems except $i,j$. In our use case, the subsystems 
 consist of $N$ spin-orbitals. Notably, the matrix is inherently symmetric, with zero values along the diagonal (by convention, as diagonal values are undefined).

The mutual information defined in Eq.~\eqref{eq:qmi} has been used in the past \cite{Huang2005, Stein2016, Ding2021} for its flexibility in describing correlation and for its further ability to discern high levels of entanglement.

In the following we will also consider the classical counterpart, the Shannon entropy, which will be useful as a comparison to the von Neumann. Shannon entropy is calculated by evolving the density matrix through a measuring channel as for the following equation:
\begin{equation}
    \label{eq:measure}
    \begin{split}
    & p(i) = \bra{i}\rho \ket{i}\\ 
    &\mathcal{M}(\rho)=\sum_{i}p(i)\ket{i}\bra{i}
    \end{split}
\end{equation}
This channel collapses the wavefunction rendering it a classical statistical distribution with probabilities $p(i)$ over the space $\ket{i}$, which in our case is the computational basis.

\section{SparQ}
\label{sec:sparq}
This section describe the procedure to retrieve efficiently quantities such as the trace and mutual information from any sparse wavefunction in qubit space.
Hereafter, since we will no longer refer to the Fock space, we will drop the $"_q"$ in the equations as we always refer to the qubit space.

For completeness, we pose the attention of the reader to the two ways of representing sparse wavefunctions. 

\begin{equation}
    \ket{\psi}=\sum_{j} \lambda_j  \left( \bigotimes_{m=1}^{N} \ket{\psi}_{j,m}\right)
    \label{eq:generalwf}
\end{equation}

The first kind of wavefunction is the one expressed by Eq.~\eqref{eq:generalwf}, which is composed of a sum of tensor products of single-qubit wavefunctions, these wavefunctions are indexed by $j$, and they sum to whichever number of components the treated wavefunction has. The second, less general representation is the binary representation \eqref{eq:binarywf}, which in previous sections and the following ones has been referred to with the index $\mathbf{i}$, and in which each $\ket{\mathbf{i}}$ is the ket of the relative binary excitation levels of the qubits.

\begin{equation}
    \ket{\psi}=\sum_{\mathbf{i}=\{0,1\}^N} k_{\mathbf{i}}\ket{\mathbf{i}}
    \label{eq:binarywf}
\end{equation}
In both notations, the main aim is to treat only a limited number of non-zero $\lambda_{j}/k_{\mathbf{i}}$.
Moreover, we notice that at this point the indexing has only counting purposes, as for Eq.~\eqref{eq:generalwf} there cannot be any particular order.

To clarify the differences between the two formalisms, we can take as an example the wavefunction on two qubits$\ket{\phi}=\left(\frac{1}{\sqrt{2}}(\ket{0}_0+\ket{1}_0)\right)\otimes \left(\frac{1}{\sqrt{2}}(\ket{0}_1+\ket{1}_1)\right)$. Clearly, by calling $\ket{+}_0=\frac{1}{\sqrt{2}}(\ket{0}_0+\ket{1}_0)$ we can rewrite $\ket{\phi}=\ket{+}_0\otimes \ket{+}_1$ following formalism \eqref{eq:generalwf}, however, we could just as well expand it in the complete binary form $\ket{\phi}=\frac{1}{2}(\ket{00}+\ket{01}+\ket{10}+\ket{11})$, which is the formalism of \eqref{eq:binarywf}. While it is the same state, it has a different description in the two notations, and also a different computational cost in its expression, both in time and in memory, as the binary form of $\ket{\phi}$ is not sparse anymore in its space of definition.
In the case of the example $\ket{\phi}$ one could resort to a local change of computational basis, which can even be done independently for each qubit, however, the division between the two formalisms remains, as this change of basis might not be done independently for each state of Eq.~\eqref{eq:generalwf}.

Beginning with the first of these two notations, we have shown in appendix \ref{appendix:mapgeneral} that any wavefunction can be encoded as Eq.~\eqref{eq:generalwf}. For fermionic wavefunctions under the Jordan-Wigner mapping, this reduces to equation \ref{eq:absorption}, with $\lambda_j$ becoming the $c_\mathbf{i}$ coefficients as in \eqref{eq:absorption}.

\subsection{Partial trace and observables}
\label{sec:operations}

The primary purpose of the procedures explained in this work is to calculate the properties of a quantum state represented as a vector in a separable Hilbert space. The measurable observables include physical quantities that directly depend on the electronic state, such as energy, dipole moment, and magnetization. 
In this work we mainly focus on the calculation of quantities that are relevant in the field of quantum information, such as the mutual information.

We can write the density matrix of the mapped state as 
\begin{equation}
    \rho=\ket{\psi}\bra{\psi}
    \label{eq:densitymatrix}
\end{equation}
that, by using Eq.~\ref{eq:generalwf}, becomes 

\begin{equation}
\begin{split}
    &\ket{\psi}\bra{\psi} = \\
    &= \sum_{j,j'} \lambda_{j}\lambda_{j'}^* \left(\bigotimes_{m=1}^{N} (\ket{\psi}_{j,m}\bra{\psi}_{j',m})\right)
\end{split}
    \label{eq:contracteddm}
\end{equation}

\subsection{Partial Trace}
\label{sub:partialtrace}
We now have the tools required to reduce the dimensionality of the qubit space starting from Eq.~\eqref{eq:contracteddm}. The reduced density matrix corresponding to the qubit $k$ can be obtained as follows:

\begin{equation}
\begin{split}
&\rho_k=tr(\rho)_{1,\cdots k-1, k+1, \cdots, N} = \sum_{j,j'} \lambda_{j}\lambda_{j'}^*  \\
& tr\left( \bigotimes_{m=1, m\neq k}^{N} (\ket{\psi}_{j,m}\bra{\psi}_{j',m})\right) \cdot \ket{\psi}_{j,k}\bra{\psi}_{j',k}
\end{split}
\label{eq:undirecttrace}
\end{equation}

From this definition, it is clear that the scaling of this operation is quadratic in the number of wavefunction components $\chi$ and linear in the number of qubits being traced $N-1$. Generally, the scaling will be $O(\chi^2N)$

\subsection{Observables}
\label{sub:observables}

Measuring observables ($O$) on the qubit space is a straightforward operation defined by:
\begin{equation}
    O = \sum_{k} o_k \bigotimes_{m=1}^{N} O_{k,m}
    \label{eq:observable}
\end{equation}
Where $O_{k,m}$ is a single qubit operator.
\begin{equation}
\begin{split}
    &\bra{\psi}O\ket{\psi} = tr(\ket{\psi}O\bra{\psi}) = \\
    &= tr \left( \sum_{j,j',k} \lambda_{j}\lambda_{j'}^*o_k \left(\bigotimes_{m=1}^{N}(\ket{\psi}_{j,m}\bra{\psi}_{j',m}O_{k,m})\right)\right)
\end{split}
    \label{eq:psiOpsi}
\end{equation}

We note that this operation has a quadratic cost in the number of states and a linear cost in the number of qubits, as described in Eq.~\eqref{eq:undirecttrace}, however, there is now also a multiplicative cost in the number of operators defining $O$.

\subsection{\label{subsec:direct}Direct tracing}

The partial trace equation introduced in \eqref{eq:undirecttrace} requires evaluating all possible combinations of two states ($\sum_{j,j'}$) in the outer product defining the density matrix~\eqref{eq:densitymatrix}, significantly limiting the number of states $\chi$ of the wavefunction~\eqref{eq:generalwf} that can be considered. Additionally, the traced density matrix produced will have the full dimension of the space, scaling exponentially. This further restricts the maximum dimension of the space that can be processed.

To overcome these limitations, we change the representation for the wavefunction using the binary representation as in Eq.~\eqref{eq:binarywf}. This approach reduces the cost of tracing from quadratic scaling in $\chi$ to linear scaling, significantly lowering the computational cost.
We begin by defining the notation, referring back to Eq.~\eqref{eq:binarywf}, where each of the $\ket{\mathbf{i}}$ represents a ket in a binary computational basis of the qubit space. To compute the partial trace, we must first define how the various subspace divisions are represented within this binary notation.

Given a ket $\ket{\mathbf{i}}$ as in Eq.~\eqref{eq:binarywf} and a subset of qubits $A$ with Hilbert space $\mathcal{H}_A$, we represent the reduction of $\ket{\mathbf{i}}$ to the space of the qubits in $A$ as $\ket{\mathbf{i}}_A$, indicating the binary collection of the qubit states with the following: 
\begin{equation}
    \left(\ket{\mathbf{i}}\right)_A = \ket{\mathbf{i}}_A
    \label{eq:red}
\end{equation}
While working with different subsets of qubits, say $A$ and $B$, we can still resort to this notation as long as the two subsets of qubits satisfy $A\cap B=\varnothing$. Furthermore, representing with $\mathcal{N}$ the set of all qubits, if $A\cup B=\mathcal{N}$, then $\ket{\mathbf{i}} = \ket{\mathbf{i}}_A \otimes \ket{\mathbf{i}}_B$.

With this notation in mind, we can now retake Eq.~\eqref{eq:partial_trace} to see how there is only an intrinsic linear cost in the definition, which is brought by the summation over all computational basis states of $B$. 

Furthermore, since we are working with a sparsely defined vector $\ket{\psi}$~\eqref{eq:binarywf} with $\chi$ components, the possible reductions to the space B are going to be at most $min(2^{|B|},\chi)$.

We can now see some differences between this approach and the one in \eqref{eq:contracteddm}, since in the latter the state did not have any intrinsic structure nor order, one could only construct the whole density matrix via outer product \eqref{eq:densitymatrix} and only then work on tracing part of the qubits.

Instead, the aim is now to start with $\ket{\psi} \in \mathcal{H}_{\mathcal{N}}$ and obtain the density matrix $\rho_A$.
\begin{equation}
\rho_{A}=tr(\ket{\psi}\bra{\psi})_{\overline{A}}
\label{eq:puretrace}
\end{equation}
To obtain a better scaling, and therefore exploiting the linearity of \eqref{eq:partial_trace}, the main idea is to consider the density matrix never doing the outer product directly, but working instead on its wavefunction expression whenever the aim is to treat a pure state. 
Moreover, we note that in Eq.~\eqref{eq:undirecttrace} the resulting traced density matrix could not always be expressed directly as it is in terms of the computational basis in $A$, or else it would take full exponential form. In this notation instead, we will obtain the density matrix directly in a sparse form linked to the computational basis of $A$. This will allow any other kind of sparse operation on the density matrix.

Elucidating the pseudo-code scheme reported in Algorithm \ref{tracecode}, the procedure consists of the following steps:

\begin{algorithm}
    \SetKwInOut{Input}{Input}
    \SetKwInOut{Output}{Output}
    \underline{Direct Trace} $(\ket{\psi}, A)$\;
    \Input{$\ket{\psi}, A$}
    \Output{$\rho_A, \mathcal{H}_A$}
    $\gamma_A \gets \{\}$\;
    \For{$\ket{\mathbf{i}} \in \ket{\psi}$}{
        \If{$\ket{\mathbf{i}}_A \notin Keys(\gamma_A)$}{
            $\gamma_A\left[\ket{\mathbf{i}}_A\right] \gets \{\}$\;
        }
        $\gamma_A\left[\ket{\mathbf{i}}_A\right] \gets \gamma_A\left[\ket{\mathbf{i}}_A\right] \cup \ket{\mathbf{i}}_{\overline{A}}$ \tcp*{Insertion of outer-space vector in the set}
    }
    $\mathcal{H}_A \gets Keys(\gamma_A)$ \tcp*{Effective Hilbert space of $A$ }
    $n_{\rho_A} \gets \#Keys(\gamma_A)$\tcp*{$\gets Dim(\mathcal{H}_A)$}
    allocate $\rho_A(n_{\rho_A},n_{\rho_A})$\tcp*{Initialized to zero}
    $c_{l} \gets 0;\quad   c_{r} \gets 0$\;
    \For{$\ket{\mathbf{a}} \in \mathcal{H}_A$}{
        $c_{r} \gets 0$\;
        \For{$\ket{\mathbf{a'}} \in \mathcal{H}_A$}{
            \For{$\ket{\overline{\mathbf{a}}} \in \gamma_A[\ket{\mathbf{a'}}]$}{
                \If{$ \ket{\overline{\mathbf{a}}} \in  \gamma_A[\ket{\mathbf{a}}]$}{ 
                $\ket{\mathbf{i}} \gets \ket{\mathbf{a}} \otimes \ket{\overline{\mathbf{a}}}; \ket{\mathbf{j}} \gets \ket{\mathbf{a'}} \otimes \ket{\overline{\mathbf{a}}}$\;
                $\rho_A(c_{l},c_{r}) \gets \rho_A(c_{l},c_{r}) + k_{\mathbf{i}}k_{\mathbf{j}}^*$\tcp*{ with $k_{\mathbf{i}}$ as in \eqref{eq:binarywf}}
                }
            }  
            $c_{r} \gets c_{r} +1$\;
        }
        $c_{l} \gets c_{l} +1$\;
    }
    \KwRet {$\rho_A, \mathcal{H}_A$}
    
\caption{Line of work for the direct tracing method.}   
\label{tracecode}
\end{algorithm}

\begin{enumerate}
    \item We start by doing an iteration over all the binary strings in $\ket{\psi}$~\eqref{eq:binarywf}(line $3$), during the iterations we save a dictionary $\gamma_A$ whose keys are all the existent binary strings in $\ket{\mathbf{i}}_A$, the items of the dictionary are a set of $\ket{\mathbf{i}}_ {\overline{A}}$. 
    
    The dictionary only saves each binary occurrence of $A$ once (line $5$), by adding another element to the $\gamma_A[\ket{\mathbf{i}}_A]$ each time there is the same occurrence of a $\ket{\mathbf{i}}_A$(line $7$). The binary strings in $A$, as keys of the dictionary, will then represent the effective space $\mathcal{H}_A$ (line $9$).
    \item After having initialized to zero the density matrix $\rho_{\mathcal{H}_A}$ to be filled, remembering Eq.~\eqref{eq:partial_trace}, we find the value of coefficient of the density matrix relative to the $\ket{\mathbf{a}}\bra{\mathbf{a'}}$, with $\ket{\mathbf{a}}, \ket{\mathbf{a'}} \in \mathcal{H}_A$, by considering all the extensions $\ket{\mathbf{\overline{a}}}$ to the traced space $\overline{A}$ in common between $\gamma_A[\ket{\mathbf{a}}_A]$ and $\gamma_A[\ket{\mathbf{a'}}_A]$, i.e.:
    $$\sum_{\ket{\mathbf{\overline{a}}} \in \gamma_A[\ket{\mathbf{a}}_A] \cup \gamma_A[\ket{\mathbf{a'}}_A] } \bra{\overline{a}}\ket{\mathbf{i}}\bra{\mathbf{j}}\ket{\overline{a}} k_{\mathbf{i}}k_{\mathbf{j}}^* $$
    With $\ket{\mathbf{i}} \gets \ket{\mathbf{a}} \otimes \ket{\overline{\mathbf{a}}}; \ket{\mathbf{j}} \gets \ket{\mathbf{a'}} \otimes \ket{\overline{\mathbf{a}}}$ (line $11-25$)
\end{enumerate}

\subsubsection{Computational Cost}
\label{subsub:directcompcost}

As shown in with the pseudo-code~\ref{tracecode}, this implementation of the partial trace involves one cycle over all the states in the superposition (line $3$) and three for-loops to define precisely $\rho_A$. However, iterating over all the $\ket{\mathbf{a'}}$ (which are in total $n_{\rho_A}$) and$\ket{\mathbf{\overline{a}}}$ entails running over all the existing combinations of $\ket{\mathbf{a'}}\otimes\ket{\mathbf{\overline{a}}}$, which are $\chi$. Therefore, the final time scaling will be $O(\chi n_{\rho_A})$. 
This scaling does not account for the tools needed to identify, save, and retrieve efficiently the correct subspaces at each step, with the most significant being the call to $\gamma_A[\ket{\mathbf{a'}}]$ and checking the common $\ket{\mathbf{\overline{a}}}$ in lines $16,17$ of \ref{tracecode}.

These operations can be executed within two different frameworks, each requiring a distinct implementation of the dictionary $\gamma_A$ and register $\gamma_A[\ket{\mathbf{a}}]$:

\begin{enumerate}
    \item Using an ordered register, each check, call, or insertion incurs a logarithmic time cost relative to the register size. In a worst-case scenario, there can be $\chi$ occurrences in the same register. Since it is reasonable for $\chi$ to scale at most polynomially in the number of qubits, checking the register $ \gamma_A[\ket{\mathbf{a}}]$ has a cost 
 scaling as $log(|\gamma_A[\ket{\mathbf{a}}]|)\sim log(N)$.
    
    The total time scaling would then be $O(\chi n_{\rho_A}log(N))$

    \item Via Hash tables; each check, call, or insertion is $O(1)$ relative to the register size, although there is an additional cost for hashing the variable $\mathbf{\overline{a}}$. This scales linearly in the number of ones in the binary string $\mathbf{\overline{a}}$, which is at most N.
    
    The total time scaling would then at most $O(\chi n_{\rho_A}*N)$
\end{enumerate}

\begin{table*}[ht]
\small
\setlength\tabcolsep{1.8pt}
\centering
\newcommand{\minitab}[2][l]{\begin{tabular}#1 #2\end{tabular}}
\begin{tabular}{|c | c | c | c| c | c |}
\hline 
Mol. &  Geometry (\AA{}) &  Basis-set & \#Qubits & tracing method & $\chi$\\ 
\hline
\hline
$H_x$ chain & \chemfig{H(-[:0,1]H)} distance, 0.745  & 6-31G & $4\cdot y$ & \minitab[c]{Quadratic\\Direct} & \\
\hline
H$_2$O & \minitab[c]{O 0.0 0.0 0.0 \\ H 0.757 0.586 0.0 \\ H -0.757 0.586 0.0}  & aug-ccpvtz~\cite{Dunning1989,Kendall1992}  & 184 & Direct & $1.2\cdot10^{5}$ \\
\hline
C$_6$H$_6$  & \minitab[c]{\chemfig{C(-[:0,1]H)} distance, 1.085\\\chemfig{C(-[:0,1]C)} distance, 1.390} & ccpvdz~\cite{Dunning1989}  & 228 & Direct& $2.5\cdot10^5$\\
\hline
\end{tabular}
\caption{Summary of tested molecular systems together with the computational details. The parameter $\chi$ is the number of \acrshort{sd}. }

\label{tab: systems}
\end{table*}

In our implementation, we resorted to the second strategyy using the particle-hole formalism explained in Appendix~\ref{appendix:particlehole}, we only make use of binary strings with a fixed number of ones relative to the number of qubits, rendering the hashing cost a constant and bringing the total cost of the operation back to $O(\chi n_{\rho_A})$. 

The memory scaling, making such heavy use of hash tables,
depends greatly on different implementations, but it can be estimated roughly as a doubling of the weight of the starting wavefunction $\ket{\psi}$ due to the copying happening in the establishment of the hash table.

This method can be generalized to partial traces of mixed density matrices which cannot be written as \eqref{eq:puretrace}, as for example a partially traced density matrix (not of a separable state), but the operation will have again a quadratic scaling in the dimension of the starting $\rho$. 

\section{Computational details\label{sec:cd}}

In the previous sections, we have introduced two main tools: encoding a sparse fermionic wavefunction into qubit space and using these wavefunctions to retrieve quantum information quantities, focusing on the partial trace of a density matrix.

We now want to present a practical analysis of our SparQ procedure to demonstrate the utility of the method. We use mutual information as the primary metric, defined by Eq.~\eqref{eq:qmi}, as it has long been an essential tool for estimating correlations in quantum systems. Its estimation, however, requires the joint density matrix of the two subsystems whose correlation is under investigation. 

The simulating framework is characterized by different packages and codes. The SparQ code implementing the trace~\eqref{eq:partial_trace} has been written in $C++$ to guarantee easy portability to parallelizing platforms such as OpenMP~\cite{Dagum1998} while the sparse wavefunctions are all relative to \acrshort{cisd} calculations found with the Pyscf~\cite{pyscf} python library.\\
Coming to the actual simulated systems, we used multiple realizations of the hydrogen chain with a 6-31G basis or the analysis of the scaling of the tracing algorithm. To test the encoding in different mappings, we analyzed the water molecule (H$_2$O) with an aug-ccpvtz~\cite{Dunning1989,Kendall1992} basis. At last, to test the full capabilities of the direct tracing algorithm, we studied how the entropy scales for different considered spaces for the benzene (C$_6$H$_6$) molecule expressed in the cc-pvdz~\cite{Dunning1989} basis-set. All the data listed here is also summarized by Tab.~\ref{tab: systems}.

\section{Results\label{sec:results}}

\begin{figure}
\centering
    a)\includegraphics[width=.50\textwidth]{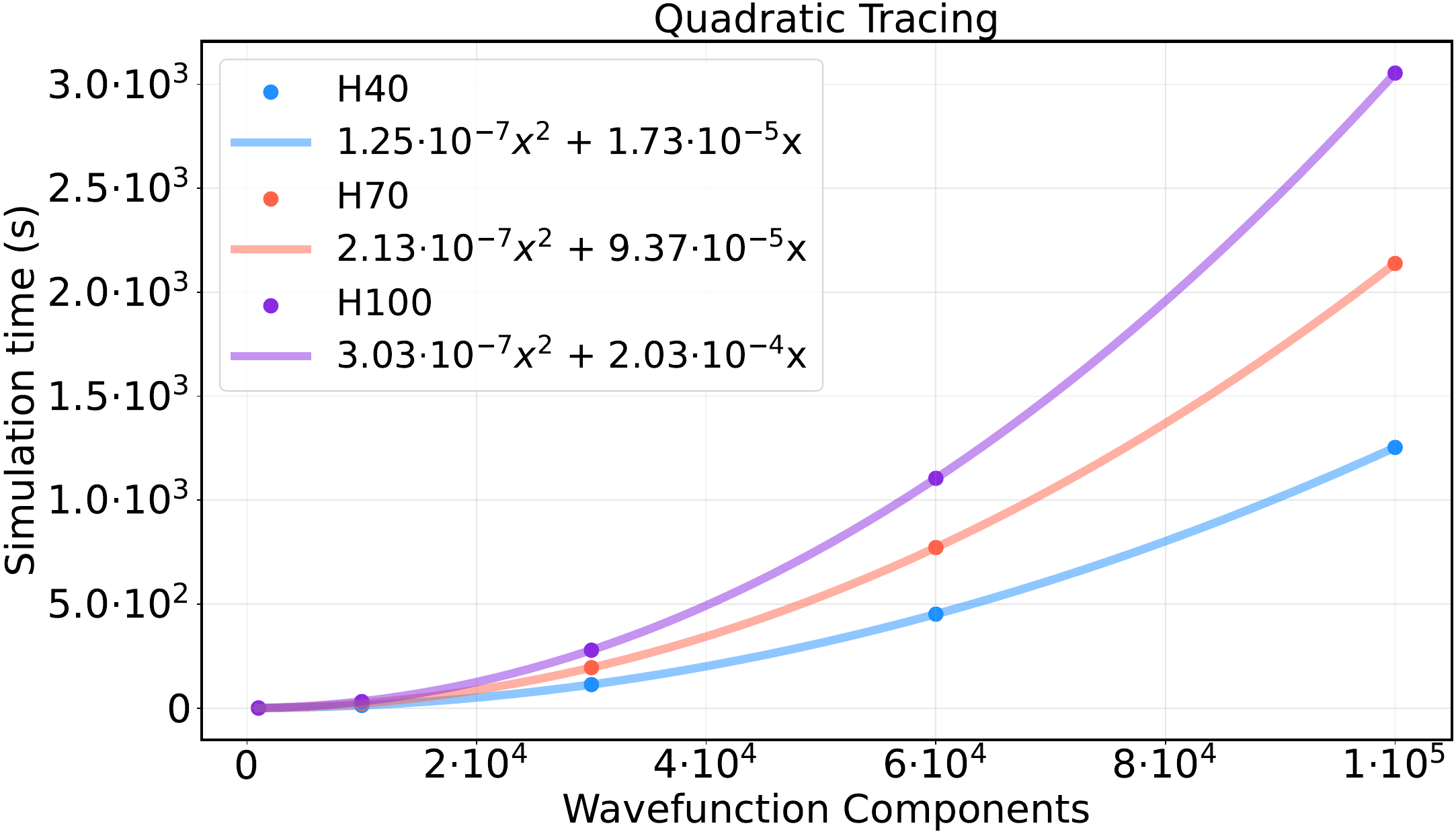}\hfill
    \\[\smallskipamount]
    b)\includegraphics[width=.50\textwidth]{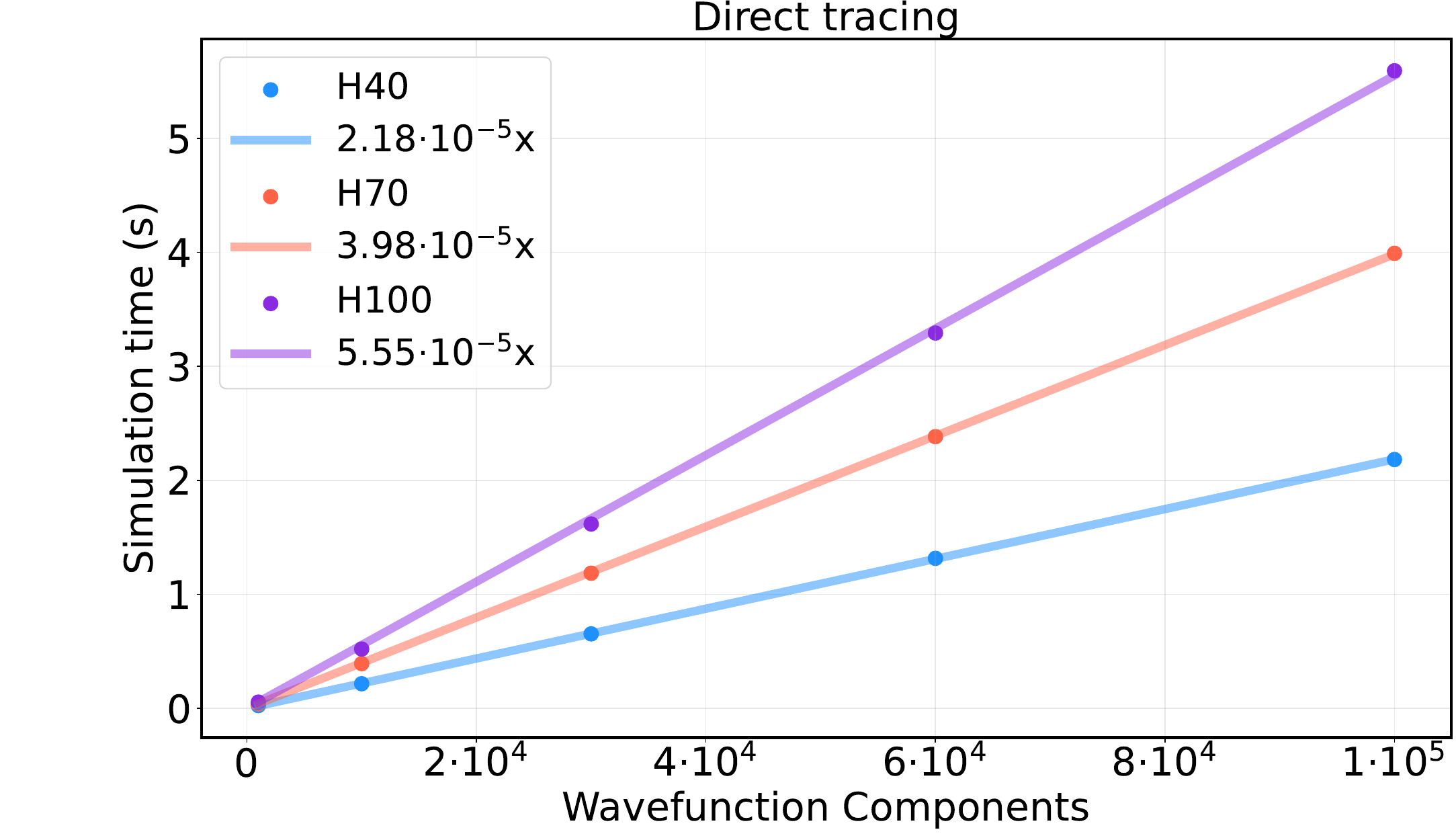}\hfill
    \caption{Time requirements for the partial trace from $N$ qubits to one qubits over different values of number of components of the wavefunction a) Quadratic tracing method of Eq.~\eqref{eq:undirecttrace}, b) Tracing method of pseudocode \ref{tracecode} with linear scaling. As shown in \ref{tab: systems} each chain $H_x$ is expressed with 4$y$ qubits.}
    \label{fig:scalingSD}
\end{figure}

\begin{figure*}[]
    a)
    \includegraphics[width=.47\textwidth]{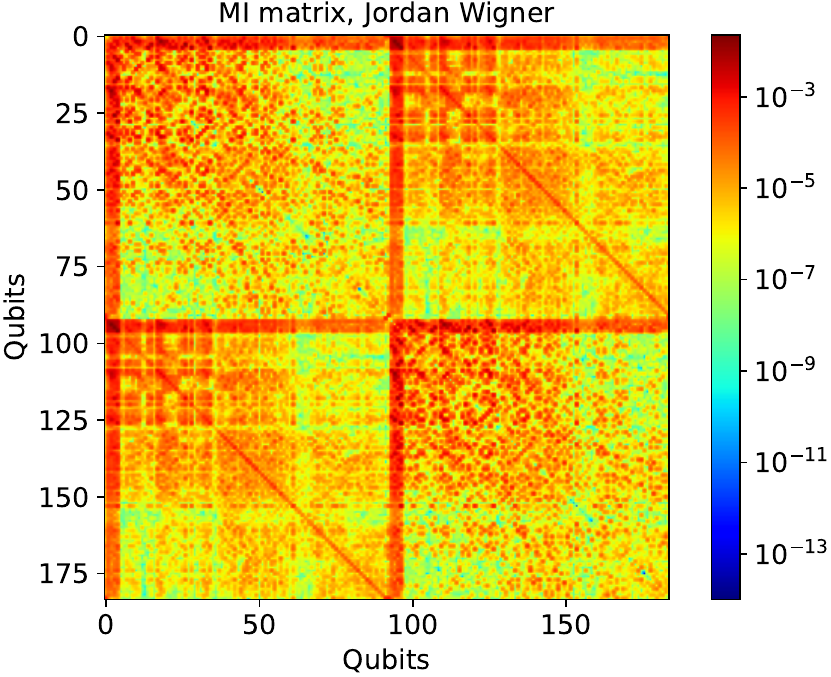}\hfill
    b)
    \includegraphics[width=.47\textwidth]{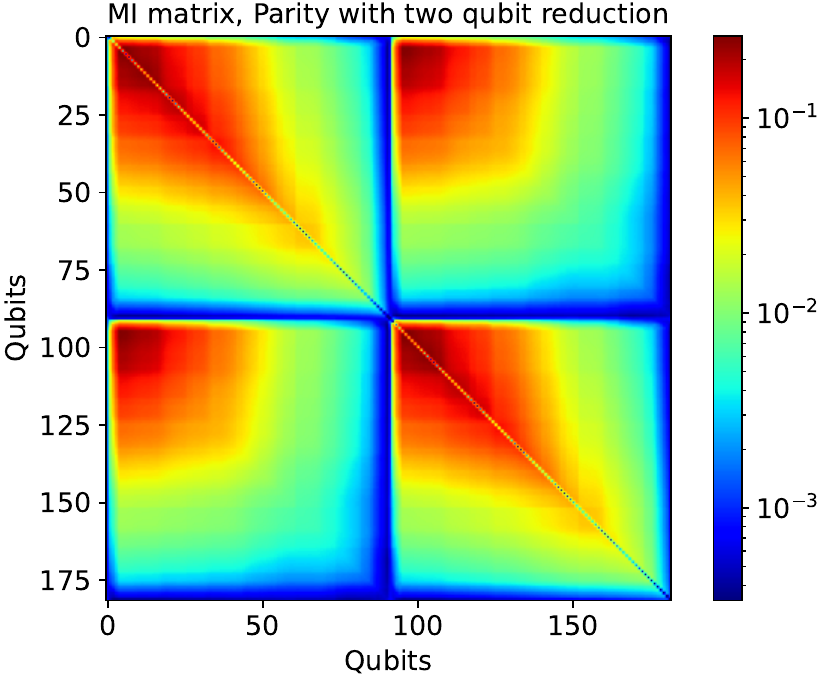}\hfill
    \caption{Mutual information for the water molecule at CISD level. Jordan Wigner mapping is reported on the left panel, parity mapping with the two-qubit reduction on the right panel. The starting wavefunction for the mutual information was the simulation for the water molecule(H$_2$O) as detailed in Tab.~\ref{tab: systems}.}
    \label{fig:h2o}
\end{figure*}
\subsection*{Scaling performance}

To verify the theoretical scaling and to evaluate the actual computational cost, we use several hydrogen chains $H_x$ of hydrogen atoms spaced $0.745$\AA{} apart to each other. Although a \acrshort{cisd} description of such systems is not appropriate, we have chosen these simple systems only for scaling tests purposes.

The results are summarized in Fig.~\ref{fig:scalingSD}, showing the CPU time required to evaluate  partial trace operation from $N$ qubits to one qubit. Image a) shows the time requirements for the quadratic scaling method described by Eq.~\eqref{eq:undirecttrace}, while image b) shows the scaling of the direct tracing method, as explained in section \ref{subsec:direct}.

Looking at image \ref{fig:scalingSD}-a), we can infer that the scaling of $O(\chi^2 N)$ for the \eqref{eq:undirecttrace} is achieved, as by fitting the parabolas not only we found a quadratic scaling for the $\chi$ parameter, but we also notice a linear scaling of the fitted coefficient of the quadratic term, coherently with the expected linear growth in $N$. This linear factor is also evident in \ref{fig:scalingSD}-b), leading to the scaling of $O(\chi*N)$, as evaluated in section~\ref{subsub:directcompcost}. This is because the shown simulation of \ref{fig:scalingSD} didn't make use of the particle-hole notation explained in appendix \ref{appendix:particlehole} specifically to show the behavior of the used Hash table.

This discussion leads us to believe that both Eq.~\eqref{eq:undirecttrace} and algorithm~\ref{tracecode} are valid methods to retrieve the partial trace and mutual information. However, while the quadratically scaling method of Eq.~\eqref{eq:undirecttrace} can hardly handle up to approximately $\sim 10^5$ wavefunction components (\ref{fig:scalingSD}-a), the latter method, maintaining a linear computational cost as shown in \ref{subsub:directcompcost}, can likely be expanded to all the available components of a wavefunction. Its limitations are primarily due to the resources required for a binary representation of the wavefunction.

\subsubsection*{Wavefunction in different mappings}

The first proof of concept involves encoding the fermionic wavefunction using different fermion-to-qubit mappings, as illustrated in Fig.~\ref{fig:h2o}. We present the mutual information matrices $I(i,j)=I_{i,j}$ of the water molecule H$_2$O using two different encoding method for the wavefunction. To retrieve these figures in the parity mapping, we used the particle-hole transformation and CNOTs permutation between mapping, explained respectively in Appendix \ref{appendix:particlehole} and \ref{appendix:gen}. 

By using Fig.~\ref{fig:h2o}, we notice that the Jordan-Wigner mapping (\ref{fig:h2o}-a) retains some chemical intuition, while the parity mapping (\ref{fig:h2o}-b) is of difficult interpretation since each qubit contains information of all the preceding ones.

\subsection*{Total entropy and active space}

\begin{figure*}[]
    \centering
    \includegraphics[width=0.8\linewidth]{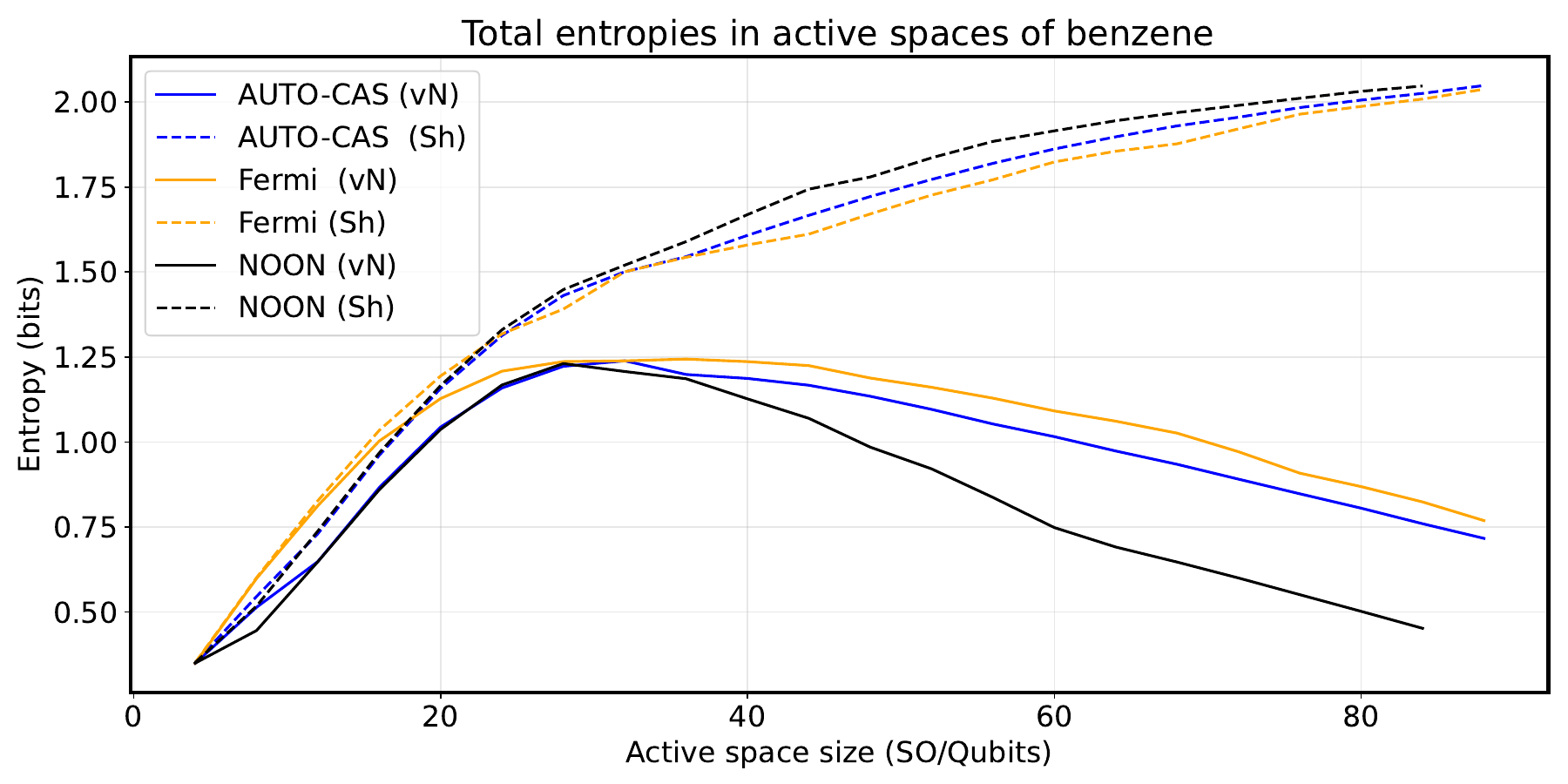}
    \caption{Total entropies as a function of the active space dimension and selection methods for a benzene molecule.The wavefunction analyzed is the \acrshort{cisd} wavefunction of benzene with cc-pvdz basis. The Natural Orbitals for the natural orbitals occupation number (NOON) method were retrieved by an iterative procedure (Iterative Natural Orbitals) also based on the \acrshort{cisd} wavefunction.
    Auto-\acrshort{cas} selects the orbitals maximizing single-orbital entropy, the Fermi method selects orbitals around the Fermi level of the \acrshort{hf} orbitals. The continuous line is the von Neumann entropy of the whole space, while the dashed line is the Shannon entropy of the whole space of the state \textit{measured} in the computational basis. Further details can be found in the main text.}
    \label{fig:active_spaces}
\end{figure*}

In the present section we would like to apply our wavefunction analysis to evaluate the total entropies of selected wavefunctions of the benzene molecule. In particular we are interested in comparing different selection methods of active spaces in \acrshort{cas}-SCF~\cite{Levine2020}. The choice of active space is not trivial, since some orbitals are more favorable than others in achieving lower energies during the orbital optimization.  Several methods have been proposed to help with this task, of which we show the following:
\begin{itemize}
    \item Fermi level selection.~\cite{Olsen2011} The active space orbital selection occurs around the Fermi level of the \acrshort{hf} occupation of the orbitals. Given its simplicity, it often misses the crucial orbitals if using the most restricted active space possible.
    \item  \acrfull{noon}~\cite{Khedkar2019}. This method analyzes the occupation number of the \acrfull{no}, which are the orbitals diagonalizing the one body reduced density matrix $\rho_{\ket{\psi},i,j}= \bra{\psi}a^\dag_i a_j\ket{\psi}$ of a given fermionic wavefunction $\ket{\psi}$. 

 The active space orbital selection is performed by including orbitals with occupation numbers that deviate the most from either empty occupation (0 electrons) or full occupation (2 electrons).\\ This method has been proven to be quite effective, since the partial occupations are an indication of the presence of correlations in a \acrshort{ci} expansion. However, since it relies on some chosen method of obtaining the wavefunction $\ket{\psi}$ and its related \acrshort{no}, it is not always applicable. Furthermore, one might also want to use different kinds of orbitals for the \acrshort{cas}-SCF, and not the \acrshort{no}.
    \item AUTO-\acrshort{cas}~\cite{Stein2016}. This method is based on an approximate \acrshort{dmrg} calculation with low bond dimension. The selection of orbitals is then based on a maximization of the single orbital entropies (see Eq.~\eqref{eq:entropy}). This method plays a similar role than the \acrshort{noon}, since an orbital entropy greater than zero also means that the orbital is neither always occupied nor always empty. However, contrary to the \acrshort{noon} selection method, it can be applied to any orbital set.
\end{itemize}

Our goal is to analyze the behavior of the total entropy of active spaces of growing size, investigating differences between the above selecting methods.

Fig.~\ref{fig:active_spaces} reports the total von Neumann and Shannon entropies as a function of number of orbitals/qubits included in the active spaces. \\ The system analyzed is the benzene molecule with a \acrshort{cisd} wavefunction which is used in all steps of the calculation.

Looking firstly at the von Neumann entropy, we can see that even if the AUTO-\acrshort{cas} selects the most entropic orbitals, the resulting entropy of the total active space is always smaller than the one of the space selected around the Fermi level. 
While this might seem counter-intuitive, it can be explained by considering that the entropy of the total untraced density matrix $\rho=\ket{\psi}\bra{\psi}$ must be zero for a pure state. Thus, the AUTO-\acrshort{cas} space is composed of the orbitals that allow to maintain, within the chosen space, the closest description to the original state $\ket{\psi}\bra{\psi}$. 

The same analysis has been performed using natural orbitals, to address the difference between the \acrshort{noon} selection method and AUTO-\acrshort{cas} in these orbitals. This comparison, however, reveals that there is a small difference, advocating for the similarity of the two methods. For this reason, in Fig.~\ref{fig:active_spaces} we only show the \acrshort{noon} for clarity (This difference is shown in the supplementary materials). An interesting property of the simulation in natural orbitals is that they allow the smallest entropy at a fixed space dimension, evidence for greater effectiveness in describing the system with a restricted space with respect to the \acrshort{hf} Canonical orbitals. This extends the already numerous studies regarding the convenient properties of the natural orbitals in quantum information~\cite{Materia2024,Ratini2024a,Aliverti2024} and quantum chemistry~\cite{Berthier1988, Illas1988}.

One last interesting perspective is obtained by looking at the Shannon entropy of the space, which, contrary to the von Neumann one, is a monotone of the space dimension, as marginalization over a probabilistic variable can only decrease classical entropy.

\section{Conclusions}
\label{sec:conclusion}
In this work have introduced SparQ, a method to efficiently calculate quantum information theoretical quantities for sparse wavefunctions in quantum chemistry. The development of \acrshort{sparq} expands over the limitations of existing methods, mostly based on Density Matrix Renormalization Group (DMRG), which struggles with chemical systems exceeding a few hundred qubits due to the non-locality of the electronic Hamiltonian. This work therefore also addresses the limitations of studies based on transition operators, by generalizing the trace operator to a wide family of fermionic-to-qubit mappings and a great number of qubits. 

By leveraging the sparsity inherent in many quantum chemistry wavefunctions, \acrshort{sparq} efficiently encodes fermionic wavefunctions into qubit space using fermion-to-qubit mappings like Jordan-Wigner. This method enables the manipulation and analysis of large-scale quantum systems, significantly extending the reach of quantum information theoretical analysis. The practical utility of \acrshort{sparq} is shown through detailed procedures for the measurement of observables in the qubit space and partial trace of a variable number of qubits scaling linearly in the number of wavefunction components.\\ 

In practical applications, \acrshort{sparq} proves valuable in estimating correlations in quantum systems, as illustrated by the mutual information analysis of the water molecule and the entropy analysis of different possible active spaces of the benzene molecule.
Overall, this demonstrated ability to handle large-scale wavefunctions opens new possibilities for understanding correlations and the behavior of complex molecular systems.

Future work will focus on refining fermion-to-qubit mappings, for example, by using the methods illustrated in the Appendices, to further reduce the computational overhead and expand the range of quantum information quantities that can be efficiently calculated using \acrshort{sparq}. The integration of various mapping techniques and the exploration of new application domains will continue to drive the evolution of quantum information analysis in Quantum Chemistry.

In summary, our work may contribute to the ongoing dialogue between quantum computing and quantum chemistry, paving the way for future innovations in hybrid quantum-classical computation and improving the understanding of correlations in quantum chemistry.

\subsection{Acknowledgments}
We are grateful to Celestino Angeli for useful discussions and for providing us with comments on the manuscript. The authors acknowledge funding from the MoQS program, founded by the European Union’s Horizon 2020 research and innovation under the Marie Skłodowska-Curie grant agreement number 955479.
The authors acknowledge funding from Ministero dell’Istruzione dell’Università e della Ricerca (PON R \& I 2014-2020).
The authors also acknowledge funding from National Centre for HPC, Big Data and Quantum Computing - PNRR Project, funded by the European Union - Next Generation EU.\\
L.G. acknowledges funding from the Ministero dell'Università e della Ricerca (MUR) under the Project PRIN 2022 number 2022W9W423 through the European Union Next Generation EU.

\newpage

\begin{appendices}
\section{Mapping generalization}
\label{appendix:mapgeneral}

\setcounter{equation}{0}
\numberwithin{equation}{section}
Even if the main focus of quantum information theory for quantum chemistry has always been within the Jordan-Wigner mapping, one might also be taken in account, for quantum computing reasons~\cite{Zhang2021, Materia2024}, different fermion-to qubit mapping. We here generalize the mapping procedure for other fermionic-to-qubit mappings.
As we already mentioned in  the Majorana operators $\gamma_{q}^{k}$ must anti-commute with each other.
Given this anticommuting relation, the Majorana pair must differ in at least one qubit, we can then assert the following:

\begin{equation}
        \mathcal{H}(a_{q,j}^{\dag})\geq 1
    \label{eq:hamming}
\end{equation}

where $\mathcal{H}$ is the Hamming distance corresponding to the qubit operator $a_{q,j}^{\dag}$ defined above.

Now, while the Jordan-Wigner\cite{Jordan1928} mapping, defined as in~\eqref{eq:jw} can be easily proven to satisfy equality in~\eqref{eq:hamming} for any pair of Majorana operators, general fermion-to-qubit mappings $\mathcal{K}$ do not have this property.
In such cases, instead of contracting and summing together the two single-qubit Pauli operators as we done in Eq.~~\eqref{eq:contraction}, we split each pairs of Majorana strings as follow
\begin{equation}
\begin{split}
    &a_{q,\mathbf{i}}^{\dag} \ket{\mathbf{0}}= \prod_{l=1}^{N} (a_{q,l}^{\dag})^{\mathbf{i}_l} \ket{\mathbf{0}} = \\
    &(a_{q,1}^{\dag})^{\mathbf{i}_1}\dots (a_{q,k}^{\dag})^{\mathbf{i}_k} \dots (a_{q,N}^{\dag})^{\mathbf{i}_N} \ket{\mathbf{0}}=\\
    &\prod_{l=1}^{N} (\frac{1}{2}\gamma_{q}^{2l-1}-\frac{i}{2}\gamma_{q}^{2l})^{\mathbf{i}_l}\ket{\mathbf{0}}=\\
    & \sum_{k=1}^{2^N}d_k\left( \bigotimes_{m=1}^{N} \sigma_{k,m} \right)\ket{\mathbf{0}}
\end{split}
\label{eq:split}
\end{equation}
where $\mathbf{i}_l=0,1$ and $\sigma_{k,m}=X_m,Y_m,Z_m,I_m$, with $d_k$ coefficients depending on the encoding method and fermionic modes appearing in $\mathbf{i}$.
Obviously, the summation runs on a number of elements that depends on the Hamming distance associated to each operator, that are no more than $2^N$. For practical cases, this number is drastically reduced. 
For example, for the Jordan-Wigner encoding method,
this expression reduce to Eq.\ref{eq:qubitSD} because the Hamming distance for each operator is equal to one.
For a generic wavefunction
\begin{equation}
    \ket{\psi}= \sum_{\mathbf{i}=\{0,1\}^N} c_{\mathbf{i}} a_{q,\mathbf{i}}^{\dag} \ket{\mathbf{0}}
\end{equation}
we obtain
\begin{equation}
   \ket{\psi}= \sum_{\mathbf{i}=\{0,1\}^N}c_{\mathbf{i}} \sum_{k=1}^{2^N}d_k  \left( \bigotimes_{m=1}^{N} \sigma_{k,m} \right) \ket{\mathbf{0}}
\end{equation}
that become
\begin{equation}
   \ket{\psi}= \sum_{j=1}^{4^N} \lambda_j  \left( \bigotimes_{m=1}^{N} \ket{\psi}_{j,m}\right) \quad \ket{\psi}_{j,m}=\epsilon_{j,m}\ket{0}_m
\end{equation}
where $\epsilon_{j,m}$ are single qubit operators and the $\lambda_j$ coefficients depends on $c_{\mathbf{i}}$ and $d_k$ ones.
We wrote that the summation runs over $4^N$ terms because this number is an upper bound but, for practical case, this is no more than polynomial: for example, for the Jordan-Wigner encoding method, we obtain Eq.\ref{eq:qubitdagtonull}, thus we have a number of strings equal to the number of Slater determinants in the wavefunction, with $\lambda_j$ being the coefficients of the Slater determinants.

On the mapped wavefunction $\ket{\psi}$, obtained from the encoding method $\mathcal{K}$, we execute the calculations explained in section \ref{sec:sparq}. It follow that the cost of these operations is proportional to the cost to encode the wavefunction, defined as the cost associated to the most expensive Slater determinant belonging to $\ket{\psi}$. So 
\begin{equation}
   C_{\mathcal{K}}(\ket{\psi})=\max_{\mathbf{i}\in \ket{\psi}} 2^{P_{\mathbf{i}}}
\end{equation}
where
\begin{equation}
    P_{\mathbf{i}}=\sum_{l=1}^N \mathbf{i}_l\Theta(\mathcal{H}(a_{q,l}^{\dag})-1)
\end{equation}
with
\begin{equation}
    \Theta(x)=
    \begin{cases}
       0 & \text{if x}  = 0 \\
       1 & \text{if x}  > 0 \\
    \end{cases}
\end{equation}
As can been easily seen, this cost is equal to 1 for the Jordan-Wigner mapping, as has been introduced above.

\section{Efficient mapping permutation}
\label{appendix:gen}
Given that fermion-to-qubit mappings are a relatively active field of research~\cite{Bravyi2002,Bravyi2017b,jkmn,Miller2023,miller2024}, we are hopeful that improvement in the understanding of mappings will bring on a further reduction in the cost of encoding wavefunctions for general mappings, either by reducing $\mathcal{C}_{\mathcal{K}}$ or by a complete revision of the method.
On this line, we can further show how there are already specific techniques to bring down such costs for specific mapping. It is widely recognized that the Parity mapping serves as the dual to the Jordan-Wigner mapping in terms of the information it encodes.

However, it can be shown~\cite{Harrison2022} that at the level of unitary transformation in the qubit space, the difference between the two can be summarized by a permutation described by the chained application of CNOTs unitary from the first to the last qubit.

\begin{equation}
    \begin{split}
            &U_{JW\rightarrow P}=CNOT_{N-1, N}\dots CNOT_{i-1, i} \dots CNOT_{0, 1}\\
            &a_{P, i}^{\dag}= U_{JW\rightarrow P}a_{JW,i}^{\dag}U_{JW\rightarrow P}^{\dag}
    \end{split}
    \label{eq:JWtoP}
\end{equation}
Considering that each piece in superposition of \eqref{eq:qubitdagtonull} for our qubit wavefunction is generated by $a_{q,\mathbf{i}}^{\dag}$, we can rewrite \begin{equation}
    \label{eq:simpleP}
    \begin{split}
        a_{P,\mathbf{i}}^{\dag} \ket{\mathbf{0}}_q = U_{JW\rightarrow P}a_{JW,\mathbf{i}}^{\dag}U_{JW\rightarrow P}^{\dag} \ket{\mathbf{0}} = \\
        = U_{JW\rightarrow P}a_{JW,\mathbf{i}}^{\dag} \ket{\mathbf{0}} = U_{JW\rightarrow P}\left( a_{JW,\mathbf{i}}^{\dag} \ket{\mathbf{0}}\right)
    \end{split}
\end{equation}
Since the zero state is an eigenstate of the CNOT and thus $U_{JW\rightarrow P}^{\dag} \ket{\mathbf{0}}=\ket{\mathbf{0}}$.
At last, since $U_{JW\rightarrow P}\left( a_{JW,\mathbf{i}}^{\dag} \ket{\mathbf{0}}\right)$ is a classical bit-wise operation, one can conclude that this operation can be efficiently done in any case at no further multiplicative cost from the Jordan Wigner mapping. Finally, we notice that since the only requirement used to establish \eqref{eq:simpleP} was the pure CNOTs composition of $U_{JW\rightarrow P}$, this procedure can be applied to any mapping in the permutation group of the starting mapping, which in this case is the Jordan Wigner mapping, further details can be found in \cite{Harrison2022}.
\end{appendices}

\begin{appendices}
\section{Particle-Hole duality}
\label{appendix:particlehole}

Up to now we encoded the fermionic vacuum state into the computational state with all the qubits in the $\ket{0}$ state, so
\begin{equation}   
\ket{\varnothing}\rightarrow\ket{\mathbf{0}}
\end{equation}
To reduce the computational effort needed to encode a wavefunction and perform some operation we take into account the particle-hole transformation. 
This then brings us to consider the Hartree-Fock state as the vacuum state and to encode it as follow

\begin{equation}
    \ket{\mathbf{HF}}\rightarrow\ket{\mathbf{0}}
    \label{eq:newnull}
\end{equation}
It follow that the operators of the fermionic modes $a_i^{\dag}$, with $i=1,\dots,N$, are transformed as follow
\begin{equation}
    a_i^{\dag}=
    \begin{cases}
     a_i^{\dag} & \text{if i} > n \\
     b_i & \text{if i} \le n \\
    \end{cases}
\end{equation}
where $b_i$ are the annihilation operators for hole particles that are defined for $i\le n$ and $n=\sum_{k=1}^N \bra{\mathbf{HF}}a_i^{\dag}a_i\ket{\mathbf{HF}}$.
Obviously, these operators obey to the Fermi algebra
\begin{equation}
    \label{eq:phcomrel}
    \begin{split}
        &\{ b_{i},b_{j}\}=\{ a^{\dag}_{k},b^{\dag}_{j}\}=\{ a_{k},b^{\dag}_{j}\}=0\\
        &\{ b_{i},b^{\dag}_{j}\}=\delta_{i,j}\mathbb{1}
    \end{split}
\end{equation}
with $i,j=1,\dots,n$ and $k=n+1,\dots,N$.
As an example, consider the following state
\begin{equation}
    a_{l}^{\dag} a_{m}^{\dag} a_{n} a_{o} \ket{\mathbf{HF}} = a_{l}^{\dag} a_{m}^{\dag} a_{n} a_{o} \prod_{i=1}^n a_{i}^{\dag} \ket{\varnothing}
    \label{eq:doubleexcitation}
\end{equation}
To map it on the qubits space, we should apply $n+4$ operators to the encoded vacuum state $\ket{\mathbf{0}}$. Taking into account the particle-hole transformation, we obtain that
\begin{equation}
    a_{l}^{\dag} a_{m}^{\dag} a_{n} a_{o} \ket{\mathbf{HF}} \rightarrow a_{q,l}^{\dag} a_{q,m}^{\dag} a_{q,n} a_{q,o}\ket{\mathbf{0}}
\end{equation}
so we just apply $4$ operators to the vacuum state. The computational cost, in this case, does not depend on the number of particles $n$ in the Hartree-Fock state, substantially reducing the overhead.

\end{appendices}

\newpage

\bibliographystyle{unsrt}
\bibliography{Cite}
\clearpage

\printglossary[type=\acronymtype]

\printglossary
\end{document}

\affil[1]{%
Dipartimento di Scienze Fisiche e Chimiche, Universit\`a degli Studi dell’Aquila, Coppito, L’Aquila, Italy }%

\affil[2]{Dipartimento di Ingegneria e Scienze dell'Informazione e Matematica\\ Universit\`a degli Studi dell'Aquila, Coppito, L'Aquila, Italy}

\affil[3]{Dipartimento di Scienze Matematiche, Fisiche e Informatiche\\ Universit\`a degli Studi di Parma, Parma, Italy}

\maketitle

%\vspace{160pt}
\section{Total entropy and active space, Natural Orbitals}
\newpage

\begin{figure}[]
    \centering
    \includegraphics[width=0.89\linewidth]{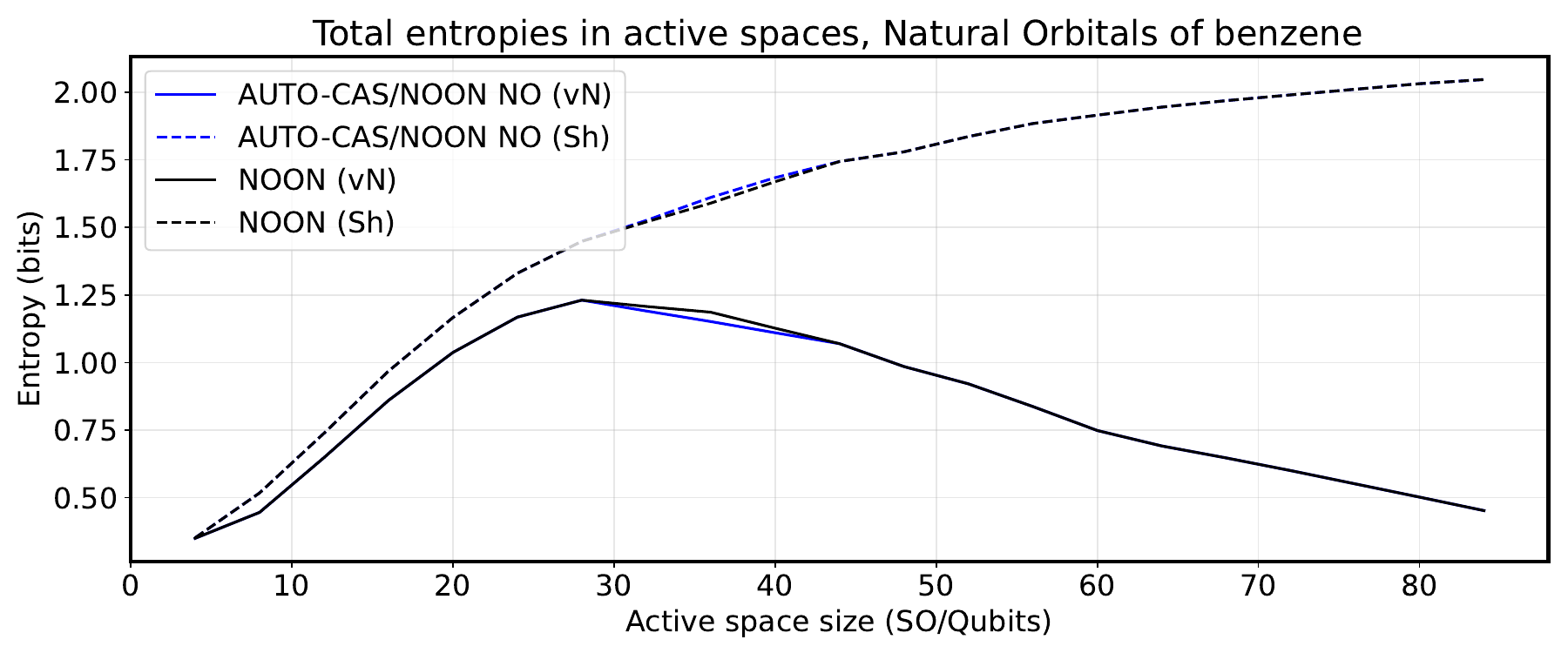}
    \caption{Entropy of the chosen active space with a growing dimension of the space with different selection methods. The wavefunction analyzed is the CISD wavefunction of the C$_6$H$_6$ molecule with cc-pvdz basis. The NO for the NOON method were retrieved by an iterative procedure (Iterative Natural Orbitals) also based on the CISD wavefunction. The continuous line is the von Neumann entropy of the whole space, while the dashed line is the Shannon entropy of the whole space of the state \textit{measured} in the computational basis.   Further details can be found in the main text.}
    \label{fig:active_spaces}
\end{figure}